\documentstyle[12pt]{article}

\def\p{\partial}
\def\fn#1{\mathop{\rm #1}}
\def\hx#1{\hat x_{#1}}

\def\hP{\hat P}
\def\fr{\frac}
\def\half{{\frac12}}

\def\1{\hbox{\bf 1}}
\def\[{\left[}
\def\]{\right]}
\def\g{\gamma}

\def\i{\item}
\def\k{\kappa}

\def\U{{\cal U}}

\def\bib{\bibitem}
\def\sec{\setcounter{equation}{0}}
\title{$\kappa$-deformed realisation of $D=4$ conformal algebra
\thanks{The paper is dedicated to the memory of Professor Jan
Rzewuski and
will appear in the memorial Issue of Acta Physica Polonica B.}
}
\author{by \\
\em  Ma{\l}gorzata Klimek \thanks{Supported by KBN grant 2P 302 087
06.} \\
 Institute of Mathematics and Computer Science,\\
 Technical University of Czestochowa,\\
 ul. Dabrowskiego 73, 42-200 Czestochowa, Poland\\ 
\\
 and\\ \em          Jerzy Lukierski \footnotemark[2] \thanks
{Partially supported by the Swiss National Science Foundation.}
\thanks{On leave of absence from the
 Institute of Theoretical Physics, University of Wroc{\l}aw,
 pl. M.\ Borna 9, 50-204 Wroc{\l}aw, Poland.} \\
Department de Physique Theorique,\\ Universite de Geneve,\\
 24 quai Ernest-Ansermet,\\ CH-1211 Geneve 4, Switzerland}
\date{}

\def\<{\left<}
\def\>{\right>}

\def\e{\epsilon}
\def\ben{\begin{enumerate}}
\def\een{\end{enumerate}}
\def\:{\,\,:\,\,\,}

\def\to#1{\,\,{\stackrel{#1}\longrightarrow}\,\,}
\def\Pc{{\cal P}_4}
\def\({\left(}
\def\){\right)}

\def\~{\widetilde}

\def\poin{Poincar\'e }

\newcounter{popnr}
\renewcommand{\theequation}{\arabic{section}.\arabic{equation}}
\def\alpheqn{\setcounter{popnr}{\value{equation}}
             \stepcounter{popnr}
             \setcounter{equation}{0}

\def\theequation{\arabic{section}.\arabic{popnr}\alph{equation}}
             }
\def\reseteqn{\setcounter{equation}{\value{popnr}}
              \def\theequation{\arabic{section}.\arabic{equation}}
              }
\newcommand{\beq}{\begin{eqnarray}}
\newcommand{\eeq}{\end{eqnarray}}
\newcommand{\beqq}{\begin{eqnarray*}}
\newcommand{\eeqq}{\end{eqnarray*}}

\def\bel#1{\begin{equation}\label{#1}}
\def\be{\begin{equation}}
\def\ee{\end{equation}}

\def\r#1{(\ref{#1})}

\def\bl{\alpheqn}
\def\el{\reseteqn}
\def\a{\alpha}
\def\b{\beta}

\def\ba{\begin{array}}
\def\ea{\end{array}}

\def\kdef{$\k$-deformed }
\def\0{^{(0)}}

\begin{document}
\begin{titlepage}
\def\thepage{}
\maketitle
\begin{abstract}
We describe the generators of $\k$-conformal transformations, leaving
invariant
the \kdef d'Alembert equation. In such a way one obtains the
conformal
extension of the off-shell spin zero realization of \kdef \poin
algebra.
Finally the algebraic structure of \kdef conformal algebra is
discussed.
\end{abstract}
\end{titlepage}
\def\thepage{\arabic{page}}

\section{Introduction}
Recently the idea of quantum deformations (see e.g. [1--3]) has been
applied
to $D=4$ \poin algbera [4--13] as well as $D=4$ conformal algebra [6,
14--16].
For the \poin algebra one can distinguish two types of deformations
\ben
\i[a)] The $q$-deformation $\U_q(\Pc)$, with the dimensionless
parameter $q$
(see e.g.\ [5]).

Because the conformal transformations describe the symmetry of the
world
without dimensionfull parameters, the existence of $q$-deformed
conformal
algebra $\U_q(O(4,2))$ should be expected. Indeed, such deformations
described by Drinfeld Jimbo scheme were proposed [6, 14--16], with
the
following inclusion valid in the algebra sector
\bel{1.1}
\U_q (\Pc) \subset \U_q(O(4,2))\,,
\ee
\i[b)] The \kdef $\U_\k(\Pc)$, with dimensionfull parameter $\k$ [4,
6--13].

In such a case the problem of finding a $\k$-deformation of the
conformal
algebra $\U_\k(O(4,2))$ such that
\bel{1.2}
\U_\k (\Pc) \subset \U_\k(O(4,2))\,,
\ee
has not been yet discussed.
\een

In this paper we would like to consider the problem of the inclusion
\r{1.2}
 on the level of particular representations. The realizations of the
\kdef
\poin algebra with arbitrary spin were given in [7, 9]. If ($\vec m$,
$\vec
l$) is any finite-dimensional realization of the standard Lorentz
algebra
($i,j,k=1,2,3$)
\bel{1.3}
\ba{rcl}
[m_i,m_j]&=& i \e_{ijk}m_k\,,\\[1mm]
[m_i,l_j]&=& i \e_{ijk}l_k\,,\\[1mm]
[l_i,l_j]&=&-i \e_{ijk}m_k\,,
\ea
\ee
then the realization on the multicomponent field $\Psi_A(p_\mu)$
($A=1\ldots
N$; $N= \fn{dim} m_i = \fn{dim} l_i$) is given by the formulae:
\bl
\bel{1.4a}
M_i= -i \e_{ijk}p_j\p_k+m_i\,,\qquad P_\mu=p_\mu\,,\\
\ee
\bel{1.4b}
N_i= i\(p_i \p_0 +\k \sinh \fr{p_0}{\k} \p_i\)+e^{\pm \frac{p_{0}}{2
\k}} l_i 
    \pm \frac1{2\k} \e_{ijk}p_jm_k\,.
\ee
\el
If $N=2S+1$ and the matrices $m_i$, $l_i$ describe the irreducible
representations $(s,0)$ and $(0,s)$ of the Lorentz algebra, we obtain
the
classical \kdef field realizations with spin $s$ (see [9]; the case
$s=\frac12$ corresponding to \kdef Dirac equation has been considered
explicitely also in [17, 18]). For the spinless case ($m_i=l_i=0$)
the boost
generators can be written as follows:
\bel{1.5}
\ba{rcl}
M_i&=& -i \e_{ijk}p_j\p_k\,,\qquad P_\mu=p_\mu\,,\\[1mm]
N_i&=& i ( p_i\hat \p_0 +\hat p_0 \p_i)\cosh \fr{p_0}{2\k}\,,
\ea
\ee
where we choose the new variables
\bel{1.6}
\hat\p_0 = \p_0\(\cosh \frac{p_0}{2\k}\)^{-1}\,,\qquad \hat p_0 =
2\k\sinh\frac{p_0}{2\k}\,,
\ee
and
\bel{1.7}
[\hat\p_0,\hat p_0 ]=1\,.
\ee
The algebra satisfied by the generators \r{1.5} is the $\k$-\poin
algebra
with the condition $\vec P\cdot \vec M = 0$, i.e.
\bl
\bel{1.8a}
[M_i,M_j] = i \e_{ijk}M_k\,,\qquad [M_i,N_j]=i\e_{ijk}N_k\,,
\ee
\bel{1.8b}
[N_i,N_j] =- i \e_{ijk} M_k\(1 +\frac{\hP_0^2}{2\k^2}\)\,,
\ee
\bel{1.8c}
[M_i,P_j]= i\e_{ijk}P_k\,,\qquad [M_i,\hat P_0]=0\,,
\ee
\bel{1.8d}
[N_i,P_j]= i \hat P_0 \(1 + \frac{\hP _0^2}{4\k^2}\)^\half
\delta_{ij}\,,
\ee
\bel{1.8e}
[N_i, \hP_0] = i P_i \(1 + \frac{\hP _0^2}{4\k^2}\)^\half\,,
\ee
\el
where we use the modified energy operator
\bel{1.9}
\hP_0 = 2 \k \sinh \frac{P_0}{2\k}
\ee
and also the supplementary condition satisfied by spinless \kdef
boost
generators:
\bel{1.10}
P_i N_j -P_j N_i = i \e_{ijk} M_k \hP_0 \(1 + \frac{\hP _0^2}{4\k^2}\)
^\half\,,
\ee
The choice \r{1.9} of the energy operator transforms the \kdef mass-
shell
condition in standard basis (see [8])
\bel{1.11}
\vec P^2 - (2\k\sinh \frac{P_0}{2\k})^2 = -M^2\,,
\ee
into the classical mass-shell condition (see also [13])
\bel{1.12}
\vec P^2 - (\hP_{0})^2 =-M^2\,.
\ee
Further in Sect.\ 2 we describe the extension of the generators
\r{1.5} of
the $\k$-\poin algebra to the generators of $\k$-conformal algebra,
by
supplementing \r{1.5} with the dilatation generator $D$ and four
conformal
generators $K_\mu$ which preserve the massless \kdef mass-shell
conditions
\r{1.11} (or equivalently \r{1.12}) with $M=0$). In such a way we
deform the
generators of the $D=4$ conformal transformations leaving invariant
classical
$D=4$ wave equation $\Box \, \varphi =0$ to the ones describing in
standard
basis [8] the invariance of the \kdef $D=4$ wave equation
\bel{1.13}
\[\Delta - \(2\k \sin \frac{\p_t}{2\k}\)^2\] \varphi (x)=0\,.
\ee

The main result of our consideration is the conclusion that the
commutators
of the generators of the \kdef conformal algebra contain not only the
nonlinearities in the fourmomentum variable, but also the quadratic
terms in
the Lorentz and dilatation generators. In Sect.\ 3 we present the
general
discussions, in particular we list some problems and give an outlook.

\sec
\section{Spinless realizations of the \kdef conformal algebra}

The classical form \r{1.12} of the \kdef mass-shell condition implies
that
the additional generators leaving invariant the massless \kdef field
equation
\r{1.13} can be written in complete analogy to the classical
conformal
generators. Denoting by
\bel{2.1}
\hx0 =x_0 \frac{1}{\cos \frac {\p_0}{2\k}}\,,\qquad D_0 = 2 \k \sin
\frac{\p_0}{2\k}\,,
\ee
the relation \r{1.7} takes the form
\bel{2.2}
[D_0,\hx0]=1\,.
\ee
If we introduce
\bel{2.3}
\vec x^2= \vec x \cdot \vec x + \hx0^2\,,
\ee
one can supplement \r{1.5} with the following generators:
\bel{2.4}
\ba{rcl}
D&=&i \vec x  \cdot \vec \p +i \hx0 D_0+ i \,,\\
K_i&=& -i\vec x^2 \p_i + 2 x_i D\,, \\
K_0&=& -i\vec x^2 D_0 + 2 \hx0 D \,.
\ea
\ee
It is interesting to observe that the algebra $O(4,1)$ formed by the
generators ($M_i$, $P_i$, $D$, $K_i$) is classical and the \kdef
conformal
generators ($N_i$, $P_0$, $K_0$) belong to the coset ${\cal K}=
\frac{O(4,2)}{O(4,1)}$. The classical algebra $O(4,1)$ can be
interpreted as
the $D=3$ Euclidean conformal algebra, extending conformally the
classical
$E_3$ subalgebra of \kdef $D=4$ \poin algebra.

We shall describe now algebraically the \kdef realizations of
conformal
algebra, given by \r{1.5} and \r{2.4}. One gets the following set of
relations:
\ben
\i[a)] The classical $O(4,1)$ algebra with the generators ($M_i$,
$P_i$, $D$,
$K_i$)
\bel{2.5}
\ba{rclrcl}
[M_i, M_j]&=& i\e_{ijk} M_k\,, & [M_i,P_j]&=& i\e_{ijk} P_k\,,\\[1mm]
[M_i,D] &=& 0\,,& [P_i,P_j]&=& 0\,, \\[1mm]
[M_i,K_j]&=& i\e_{ijk}K_k\,,& [D,P_i]&=&-i P_i\,,\\[1mm]
[D,K_i]&=& i K_i\,,& [K_i\,K_j]&=&0\,,\\[1mm]
[P_i,K_j]&=& -2i \delta_{ij}D - 2i \e_{ijk}M_k\,.
\ea
\ee
\i[b)] The $O(4,1)$ covariance relations of the coset generators
($N_i$,
$K_0$, $\hP_0$):
\def\suma{{1+\frac{\hP_0^2}{4\k^2}}}
\bel{2.6}
\ba{rcl}
[M_i, N_j ]&=& i \e_{ijk} N_k\,,\qquad [M_i, \hP_0] = [M_i,  K_0 ] =
0\\[1mm]
[P_i,N_j]&=& -i \delta_{ij} \hP_{0} \(\suma\)^\half\,,\quad
[P_i,\hP_0]=0\\[1mm]
[P_i, K_0] &=& -2 i N_i\(\suma\)^{-\half}\,,\\[1mm]
[D,N_i]&=& - N_i \frac {\hP_0^2}{4\k^2}\(\suma\)^{-1}\,,\\[1mm]
[D,\hP_0]&=& - i \hP_0 \,, \qquad [D,K_0]=iK_0\,,\\[1mm]
[K_i,N_j]&=& -i\delta_{ij} K_0 \( \suma\)^\half - i N_j N_i
\frac{\hP_0}{2\k^2}
\(\suma\) ^{-\frac32}\\
&& + \frac{1} {4\k^2} N_j P_i \(1-\frac{\hP_0^2}{2 \k^2} \)
\(\suma\)^{-2} \\[1mm]
[K_i,\hP_0]&=& -2i N_i \( \suma \)^{-\frac12} \,.
\ea
\ee
\i[c)] The relations for the coset generators
\bel{2.7}
\ba{rcl}
[N_i, N_j]&=& -i \e_{ijk} M_k\(1+\frac{\hP_0^2}{2 \k^2} \)\,,\\[1mm]
[N_i, \hP_0]&=& - i P_i \(\suma\)^\half\,,\\[1mm]
[N_i, K_0] &=& i K_i \(\suma\)^{\half} + i N_i D \frac{\hP_0}{2\k^2}
                     \(\suma\)^{-1}\\
         & & - \frac{1}{4 \k^2} N_i \hP_0 \(1-\frac{\hP_0^2}{2 \k^2}
\)
         \( \suma\)^{-2}\,,\\[1mm]
[\hP_0,K_0]&=& 2iD\,.
\ea
\ee
\een
We see from the relations \r{2.6}--\r{2.7} that the commutators of
the boosts
$N_i$ with the conformal generators $K_\mu=(K_i,K_0)$ contains the
bilinear
terms in boosts and the dilatation generators. One can conclude
therefore
that the algebra with \r{2.5}--\r{2.7} describes the quadratic
algebra with
the energy--dependent structure constants, which in the limit $\k
\rightarrow
\infty$ provides the $D=4$ classical conformal algebra.

\sec
\section{Discussion and outlook}
In the presented paper we consider the $\k$-deformation of the
spinless
representatons of the $D=4$ conformal algebra. This is only the first
step in
the programme of description of $\k$-deformation of $D=4$ conformal
algebra.
The next step consists in description of the representations of the
$D=4$
conformal algebra with arbitrary spin, containing in the $\k$-\poin
sector
the realization \r{1.4a}--\r{1.4b}.  At present we know only how to
add to
\r{1.4a}--\r{1.4b} the  \kdef dilatation generator; the form of the
\kdef
conformal generators $K_\mu$ for arbitrary spin are not known yet.
The
$\k$-deformation of the conformal representations with arbitrary spin
would
permit to generalize the algebra \r{2.5}--\r{2.7}, without the
condition
$\vec P \cdot \vec M = 0$ and \r{1.10}.

Let us recall that the \kdef \poin algebra takes the form:
\bel{3.1}
\ba{rcl}
[M_{\mu\nu},M_{\rho\tau}]&=& f^{(\k)}_{\mu\nu,\rho\tau}{}^{\a\b}(\vec
P,
P_0)M_{\a\b}\,,\\[1mm]
[M_{\mu\nu}, P_\rho]&=& f^{(\k)}_{\mu\nu,\rho}{}^\tau(P_0)
P_\tau\,,\\[1mm]
[P_\mu,P_\nu]&=&0\,,
\ea
\ee
where the ``soft'' structure constants can be easily reproduced from
the
formulae for the $\k$-\poin algebra.

The spinless case is characterized by the lack of dependence of the
``soft''
structure constant $f^{(\k)}_{\mu\nu,\rho\tau}{}^{\a\b}$ on the
three-momentum $\vec P$, i.e.
\bel{3.2}
f^{(\k)}_{\mu\nu,\rho\tau}{}^{\a\b}(\vec P, P_0)
\to{\rm spin~0} f^{(\k)}_{\mu\nu,\rho\tau}{}^{\a\b}(P_0)\,.
\ee

In the case of $D=4$ conformal algebra the modification due to the
$\k$-deformation is stronger --- instead of inhomogeneous ``soft''
Lie algebra\footnote
{The notion of ``soft'' Lie algebra has been introduced in [19,
20] and used extensively in the supergroup manifold approach to
supergravity.}
one obtains the quadratic algebra with the fourmomentum-dependent
structure
constants. Denoting by $H_i$ the $O(4,1)$ generators with the algebra
given
by \r{2.5} and by $K_\a$ the $\frac{O(4,2)}{O(4,1)}$ coset generators
(see
\r{2.6}--\r{2.7}) the algebraic relations can be written as follows
\bel{3.3}
\ba{rcl}
[H_i,H_j]&=& c_{ij}^k H_k\,,\\[1mm]
[H_i, K_\a]&=& c^{(\k)}{}_{i\a}{}^\b (P_0) K_\b +
c^{(\k)}{}_{i\a}{}^{\beta \gamma}(P_0) K_{\beta}
K_{\gamma} + c^{(\k)}{}_{i \alpha}^{\beta j} (P_0) K_{\beta} H_{j}
\,,\\[1mm]
[K_\a,K_\b]&=& c^{(\k)}{}_{\a\b}^i (P_0) H_i +
c^{(\k)}{}_{\a\b}^\g(P_0)
K_\g + c^{(\k)}{}_{\a\b}^{\g i} (P_0) K_\g H_i\,.
\ea
\ee
We expect that in the case with noncommuting spin at least some
``soft''
structure constants in \r{3.3} might depend on the three-momenta
$P_i$.

Having only scalar representations of $\k$-conformal algebra at
present we
are not able to describe its tensor products, which involves the
representations with any spin. In other words, it is now too early
too
discuss the Hopf algebra structure of \kdef conformal algebra. The
question
whether the Hopf algebra structure of \kdef conformal algebra exists
is an
open problem. In the case of positive ansver it would be interesting
to put
the \kdef algebraic structure described by the relations \r{3.3} into
the
framework of bicrossproduct  Hopf algebra [21, 22].

\def\bib{\bibitem}

\end{document}